# HARNESSING CUSTOMIZED BUILT-IN ELEMENTS: EMPOWERING COMPONENT-BASED SOFTWARE ENGINEERING AND DESIGN SYSTEMS WITH HTML5 WEB COMPONENTS

Hardik Shah

Department of Information Technology, Rochester Institute of Technology, Rochester, New York, USA

## ABSTRACT

*Customized built-in elements in HTML5 significantly transform web development. These elements enable developers to create unique HTML components tailored with specific design and purpose. Customized built-in elements enable developers to address the unique needs of web applications more quickly, supporting consistent user interfaces and experiences across diverse digital platforms. This study investigates the role of these features in Component-Based Software Engineering (CBSE) and Design Systems, emphasizing the benefits of code modularity, reusability, and scalability in web development. Customized built-in elements enable developers to address the unique needs of web applications more quickly, supporting consistent user interfaces and experiences across diverse digital platforms. The paper also discusses the difficulties and concerns that must be addressed when creating customized built-in elements, such as browser compatibility, performance optimization, accessibility, security, styling, and interoperability. It emphasizes the importance of standardization, developer tooling, and community interaction in order to fully realize the potential of these features. Looking ahead, customized built-in elements have potential in a variety of applications, including the Internet of Things (IoT), e-commerce, and educational technologies. Their incorporation into Progressive Web Apps (PWAs) is expected to further improve web experiences. While obstacles remain, the article concludes that HTML5 customized built-in elements are a driver for web development innovation, allowing the production of efficient, adaptive, and user-centric web applications in an ever-changing digital context.*

## 1. INTRODUCTION

Web development is at the forefront of innovation and user-centric design in today's quickly expanding digital landscape. The advent of HTML5 customized built-in elements marks a watershed moment in the realm of web development. These elements, a subset of the Web Components standard, have the potential to transform software engineering, Design Systems, and user experience. Their ability to contain exact functionality, visual aesthetics, and interactive behaviors represents a new paradigm in web development, encouraging code modularity, reusability, and scalability. Customized built-in elements represent a significant change in the world of web development. They allow developers to construct their own HTML components with unique features and behaviors that are similar to typical HTML elements such as <div> or <p> [6]. This adaptability is key to the ideas of Component-Based Software Engineering (CBSE), which enable developers to divide complicated applications into modular, granular components [10]. These components are intended to fit easily into the larger software ecosystem, promoting modularity, reusability, and maintainability.

Additionally, customized built-in elements have a revolutionary impact on Design Systems, which are critical in guaranteeing visual consistency and user experience across digital applications. These elements serve as the foundation for constructing user interface components by carefully adapting them to match an application's specific design language and brand identity. This personalization guarantees a consistent and visually appealing user interface across a wide range of digital products and platforms, improving brand coherence [9]. Cross-browser compatibility, speed optimization, accessibility, stylistic methods, and security are all factors to consider when implementing customized built-in elements. Attention to these details is critical for efficient integration into current online applications. In the future, they are set to revolutionize the field of web development across a diverse array of applications, including the Internet of Things (IoT), e-commerce platforms, and educational tech solutions [21]. Their interaction with Progressive Web Apps (PWAs) has the potential to improve web experiences even further. To realize their full potential, issues such as browser standardization, accessibility, security, developer tooling, and community participation must be addressed.

## 1.1. Objectives and Purpose

The following are the objectives and purpose of this paper:

Objectives:
- Investigate the importance of customized built-in elements in HTML5 Web Components.
- Emphasize their importance in Component-Based Software Engineering (CBSE) and Design Systems.
- Highlight the advantages that code modularity, reusability, and scalability provide in web development.
- Promote the broad usage of customized built-in elements to alter software engineering, improve design coherence, and improve user experience.

Purpose:
This paper's goal is to shed light on the transformative potential of customized built-in elements in current web development. When used successfully, these aspects have the potential to change software engineering and Design Systems. They improve code modularity, reusability, and scalability while adhering to CBSE principles. Furthermore, they ensure consistency and customization in the field of Design Systems, resulting in a unified and visually appealing user interface. This paper attempts to raise web development by addressing the expectations for efficiency and adaptability in digital solutions by pushing for the adoption of these aspects.

## 2. WEB COMPONENTS API AND CUSTOM ELEMENTS

## 2.1. Custom Elements in HTML5 Web Components

HTML5 Web Components feature a dynamic capability enabling developers to craft and establish their unique HTML elements, referred to as custom elements. These custom components provide a distinct and robust feature, allowing developers to encapsulate functionality, structure, and style similar to their well-known HTML equivalents, such as <div>

and <p>. Custom elements, at the heart of the Web Components standard, serve as a gateway to increasing HTML's vocabulary [4]. This enhancement streamlines the development of reusable and modular web application components [3]. The benefit of custom elements is their adaptability, which allows developers to easily define these elements using JavaScript . Once defined, these custom elements integrate smoothly into web pages, behaving exactly like any other HTML element [6].

```js
// Create a class for the element
class MyCustomElement extends HTMLElement {
  static observedAttributes = ["color", "size"];

  constructor() {
    // Always call super first in constructor
    super();
  }

  connectedCallback() {
    console.log("Custom element added to page.");
  }

  disconnectedCallback() {
    console.log("Custom element removed from page.");
  }

  adoptedCallback() {
    console.log("Custom element moved to new page.");
  }

  attributeChangedCallback(name, oldValue, newValue) {
    console.log(`Attribute ${name} has changed.`);
  }
}

customElements.define("my-custom-element", MyCustomElement);
```

Figure 1: Example of MyCustomElement and lifecycle events using HTML5 custom elements API.
Source: Adapted from [7]

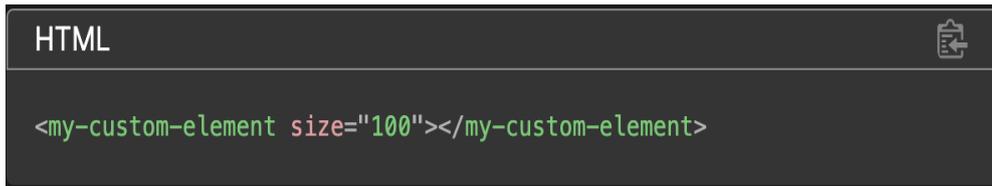

Figure 2: HTML declaration of MyCustomElement. Source: Adapted from [7]

In this case, we'll make a custom element called <my-custom-element />. This new custom element is defined in runtime using JavaScript. We established a class MyCustomElement that extends HTMLElement. If required, we can use attachShadow method to generate a shadow DOM within its constructor. We build a <my-custom-element /> element and define its content and styling. Several lifecycle methods required to get access to the component are available as part of the API - connectedCallback, disconnectedCallback, adoptedCallback and attributeChangedCallback. Following that, the customElements.define method is used to register the custom element "my-custom-element" for use in HTML. After you've defined it, you can use <my-custom-element></my-custom-element> like any other HTML element. If you pass attributes - size and color, any change to these attribute values will trigger attributeChangedCallback and eventually log the attribute name in the browser console. This example implementation highlights the power of HTML5 Web Components' custom elements, which allow you to construct reusable and encapsulated components with their own behavior and appearance.

## 2.2. Customized Built-In Elements: Tailoring Web Components for Specific Needs

A distinct and highly specialized class of entities known as customized built-in elements develops from the vast geography of the Web Components API. They derive its semantic meaning from the base element which it is extending. These custom elements are painstakingly developed to meet individual web applications' precise and frequently sophisticated needs [8]. Developers have incredible control and accuracy when creating and improving these elements, adapting them to encompass precise functionality, visual aesthetics, and interactive behaviors [1]. Customized built-in elements are distinguished by their extensive feature sets, which include a wide range of properties, methods, and event-driven mechanisms. Each component in this system has been meticulously crafted to integrate seamlessly with the distinct architecture of a particular application [8]. This seamless integration is a cornerstone, increasing code modularity and reusability by enclosing complicated and multifarious functions into self-contained components [2].

In this example, we'll be creating a customized built-in element named `plastic-button`, which behaves like a normal button but gets fancy animation effects added whenever you click on it. We start by defining a class, just like before, although this time we extend `HTMLButtonElement` instead of `HTMLElement`:

```
class PlasticButton extends HTMLButtonElement {
  constructor() {
    super();

    this.addEventListener("click", () => {
      // Draw some fancy animation effects!
    });
  }
}
```

When defining our custom element, we have to also specify the `extends` option:

```
customElements.define("plastic-button", PlasticButton, { extends: "button" });
```

In general, the name of the element being extended cannot be determined simply by looking at what element interface it extends, as many elements share the same interface (such as `q` and `blockquote` both sharing `HTMLQuoteElement`).

To construct our customized built-in element from parsed HTML source text, we use the `is` attribute on a `button` element:

```
<button is="plastic-button">Click Me!</button>
```

Figure 3: Code example for Customized built-in element. Source: Adapted from [9]

## 3. ADVANTAGES OF CUSTOMIZED BUILT-IN ELEMENTS

HTML5's custom elements feature allows developers to specify and create their own HTML elements. These custom components, like built-in HTML elements (e.g., <div>, <p>), can encapsulate functionality, structure, and styling and are a cornerstone of the Web Components standard. They offer a robust method to expand the lexicon of HTML, simplifying the process of developing modular and reusable components for web applications. The beauty of custom elements is their JavaScript-based definition, allowing them seamlessly integrate into web pages like any other HTML element. The Web Components API adds a new category of customized built-in elements, representing a significant advancement in web development. These custom elements are painstakingly developed to meet individual web applications' distinct and nuanced needs [10]. Developers possess the distinct ability to design and precisely adjust these components.

Customized built-in elements are defined by their adaptability, frequently spanning a diverse set of attributes, methods, and events, all precisely crafted to integrate seamlessly with the architecture of a particular application [5]. These components act as foundational elements that promote reusability and the modular structure of code. They achieve this by encapsulating complex and specific functionalities within self-sufficient, readily deployable units. This transformational approach to web development improves software development productivity and develops a culture of modular, manageable, and scalable code [10].

Figure 4: Word count component implemented as a Customized built-in element.
Source: Adapted from [11]

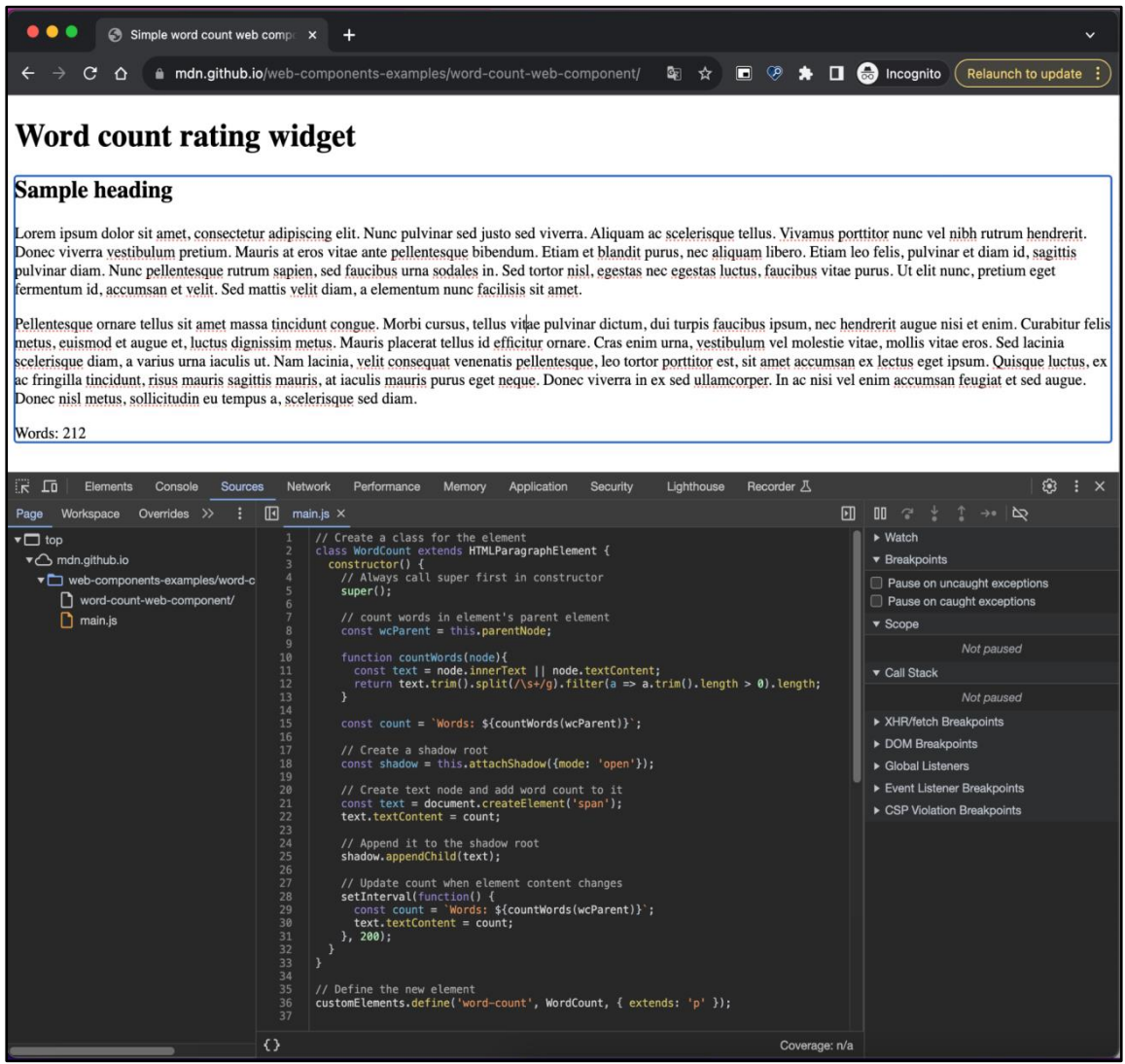

Figure 5: Source code for Word count component. Source: Adapted from [11]

Customized built-in elements are the foundations that allow developers to navigate the complex landscape of web application development with precision and grace, providing bespoke solutions to match the particular demands of any digital venture. For instance, <p is="word-count" /> element can provide a consistent and feature-rich paragraph component with a word count feature which is consistent across different web applications, reducing the need for third-party libraries and ensuring a uniform look and feel. Similarly, a  element implemented using customized built-in elements API can automatically fetch and display a user profile picture from a defined source, simplifying the process of user interface development. To conclude, customized built-in elements give developers the accuracy and flexibility they need to handle the unique needs of their digital projects, making web development more efficient and adaptable to a wide range of applications and industries.

# 4. APPLICATIONS OF CUSTOMIZED BUILT-IN CUSTOM ELEMENTS

## 4.1. Elevating Component-Based Software Engineering (CBSE)

Introducing customized built-in elements is disruptive in Component-Based Software Engineering (CBSE). These aspects are essential in systematically creating software components, allowing developers to break complex and multifaceted applications into more manageable, granular entities. Each of these components has been painstakingly designed to interact with the overall architecture and requirements of the software system [3]. Customized built-in elements prove to be the cornerstone in this process, infusing CBSE with a profound feeling of modularity, reusability, and maintainability [12]. Developers start on a quest to optimize the development process to unparalleled levels of efficiency by carefully utilizing these factors. The intrinsic compatibility of customized built-in parts with web application core architecture ensures that software components coexist healthily within the larger software ecosystem [6].

In real-world CBSE projects, customized built-in elements can be used to create complex data grids and interactive dashboards that are central to enterprise applications. Consider the case of an E-commerce company which has several dashboards in their online public shopping web application to represent latest sales and recent orders for the customer. Building new dashboards from scratch would come at an extremely high effort and cost. Customized built-in elements can address these challenges by allowing reusability of existing components in the Dashboards by migrating to customized built-in elements to provide a seamless and interactive user experience for data manipulation. However, developers face challenges such as ensuring the performance and security of these components. Custom elements help address these challenges by allowing encapsulation of functionality, which can lead to performance optimizations and better security through shadow DOM.

## 4.2. Design Systems Empowered by Customized Built-In Elements

The adaptability of customized built-in pieces extends much beyond the limitations of Component-Based Software Engineering (CBSE). These aspects emerge as vital instruments with transformative potential within the vast area of web-based Design Systems [5]. Design frameworks, serving as repositories for recyclable web UI components, design principles, and established guidelines, are crucial in maintaining uniformity and consistency across various digital platforms and brand environments, thereby supporting design and user experience [4]. Within this framework, customized built-in elements stand tall as the foundation upon which many UI components manifest. These elements are used to precisely build buttons, input fields, navigation bars, and critical interface elements. What distinguishes them is the artistry of customization, which perfectly aligns each aspect with the application's distinctive design language and brand identity. This thorough alignment is the key to developing a coherent and visually appealing user interface that connects with the brand's character.

Organizations are prepared to start on a journey of frictionless consistency by seamlessly incorporating customized built-in elements into Design Systems [5]. They may easily transmit a consistent and visually pleasant user interface throughout the varied spectrum of their digital products and platforms using these elements as their base [12]. Consequently, individuals can traverse online environments experiencing both comfort and aesthetic appeal, due to the inventive brilliance of customized built-in elements. This marks the dawn of a novel phase in the evolution of Design Systems development. [3].

Consider a huge e-commerce company with many digital channels, such as a website, mobile app, and even voice-activated purchasing assistants. The company decides to establish a complete design system in order to maintain a uniform and visually appealing user experience across all of these platforms. This design system has components in the component library based on HTML5 customized built-in elements. The corporation can migrate their existing buttons in their application to the customized built-in element called 'SignUp' by adding just one attribute to their existing HTML5 buttons and importing the new design system library. This SignUp button has been precisely crafted to complement the company's brand identity and design language. It includes variables such as size, color, and shape, allowing developers to adjust button look based on platform constraints while maintaining brand consistency.

Here's how this example relates to Design Systems empowered by customized built-in elements:

- Consistency: The e-commerce company guarantees that buttons throughout its website, app, and voice-activated assistants have a uniform look and feel by leveraging customized built-in elements like <button is="SignUp">Sign Up</button>.
- Alignment with Brand Identity: Using the customization options in <button is="SignUp">, the organization may align each button with its individual brand identity, resulting in a consistent and visually appealing user interface.

Customized built-in elements, such as <button is="SignUp">, serve as modular building blocks within the design system. The same API can be used to generate a variety of UI components such as input fields, navigation bars, and other elements, enabling reusability and efficient design revisions. As a result, customized built-in elements enable the organization to create a unified design system that assures a coherent and visually appealing user experience across varied digital products and platforms while accommodating platform-specific requirements.

## 5. CONSIDERATIONS IN IMPLEMENTING CUSTOMIZED BUILT-IN ELEMENTS

To effectively utilize HTML5 customized built-in elements, one must possess an in-depth knowledge of the technical intricacies and design elements involved in web development. These elements, while powerful, come with a set of considerations that developers must address to ensure their effective integration into web applications. Firstly, browser compatibility is a primary concern. While modern browsers have embraced the Web Components standard, discrepancies remain in how different browsers handle custom elements, necessitating the use of polyfills for unsupported features [13]. Developers must test their custom elements across a spectrum of browsers to guarantee consistent behavior and appearance [6].

Performance optimization is another critical consideration. Custom elements can introduce performance bottlenecks, particularly if they contain complex logic or are used extensively on a page. Developers should measure the impact of their elements on page load times and runtime performance, optimizing through techniques such as lazy loading and avoiding excessive DOM manipulation. Accessibility is a non-negotiable aspect of web development. Designing customized built-in elements should always consider accessibility, making sure they are functional and accessible for individuals with disabilities. This includes semantic structure, keyboard navigability, and ARIA roles where appropriate.

Styling customized elements requires a strategy that balances encapsulation with flexibility. While Shadow DOM provides style encapsulation, developers must also provide a means for consumers of the element to customize styles as needed, often through CSS custom properties or slots. State management within custom elements must be handled with care to avoid tightly coupling the elements to a specific state management solution. Instead, elements should expose a clear API for state updates and changes. Security considerations are paramount, as custom elements can be susceptible to the same range of vulnerabilities as any web technology. Developers must sanitize content to prevent cross-site scripting (XSS) and ensure that any data bindings are secure.

Interoperability with other web components and frameworks is essential. Custom elements should be designed to work within different contexts and alongside other components, which may involve managing events and data flow between components. Testing customized built-in elements is as important as testing any other part of the application. Automated testing should cover the functionality of the element, its response to state changes, and its behavior under different conditions. Documentation is often overlooked but is critical for the adoption and maintenance of custom elements. Comprehensive documentation should cover the API, usage examples, and any quirks or limitations [5]. Lastly, lifecycle management is a technical consideration where developers must handle the creation, connection, disconnection, and attribute changes of custom elements with lifecycle callbacks provided by the Web Components API [12]. In summary, the implementation of HTML5 customized built-in elements requires careful consideration of cross-browser compatibility, performance, accessibility, styling, state management, security, interoperability, testing, documentation, and lifecycle management. Addressing these considerations is crucial for the successful integration of custom elements into modern web applications.

## 6. FUTURE OF CUSTOMIZED BUILT-IN CUSTOM ELEMENTS

The horizon for customized built-in elements is expansive and promising, with the potential to revolutionize web development in profound ways. As we look to the future, several applications and challenges come into focus, heralding a new era of innovation and user-centric design. Emerging Applications: The future applications of customized built-in elements are diverse. Within the domain of the Internet of Things (IoT), these components act as the medium for intricate interactions between devices, facilitating user-friendly control interfaces and enhanced visualization of data [13]. In e-commerce, customized elements can provide unique shopping experiences with interactive and personalized components that enhance user engagement [14]. Educational technology can leverage these tools to develop interactive and adaptive learning settings tailored to each student's unique requirements [15].

Integration with Progressive Web Apps (PWAs): Customized built-in elements are set to play a significant role in the development of Progressive Web Apps (PWAs). They can be used to create app-like experiences within the browser, complete with offline capabilities and device-specific integrations [16]. This synergy will likely drive further adoption of PWAs as businesses seek to provide seamless experiences on both desktop and mobile [9].

## 7. CHALLENGES AND CONSIDERATIONS

Despite the potential, there are challenges that need to be addressed. One of the primary concerns is the standardization across browsers. While major browsers support custom elements, there are inconsistencies in implementation that can lead to compatibility issues [17]. Performance optimization is another challenge, as the complexity of custom elements can impact load times and runtime efficiency [18].

Enhancing Accessibility: Accessibility will remain a critical challenge. When creating customized built-in components, it's essential to prioritize accessibility from the beginning. These components should adhere to the Web Content Accessibility Guidelines (WCAG) to ensure they are accessible and user-friendly for individuals with disabilities [19].

Security Implications: Security is another area of concern. Custom elements that handle data must be designed to prevent vulnerabilities such as cross-site scripting (XSS) and ensure data privacy [20].

Tooling and Developer Experience: Advancing the design of more advanced tools will be crucial for facilitating the creation and upkeep of customized built-in elements [5]. Integrated development environments (IDEs) and frameworks will need to evolve to provide better support for debugging and testing these components [21].

Standardization and Community Engagement: The evolution of web standards will continue to shape the future of customized built-in elements. Active engagement with the web standards community will be crucial to ensure that the development of custom elements aligns with the evolving needs of the web [22].

In conclusion, the future of customized built-in elements is bright but requires careful navigation of emerging technologies, standards, and user expectations. As the web continues to evolve, these elements will be at the forefront of creating more dynamic, efficient, and user-friendly web applications.

## 8. CONCLUSION

HTML5 customized built-in elements represent a pivotal advancement in the realm of web development. They provide developers with a robust mechanism for crafting bespoke HTML components that possess specific functions and aesthetics. This approach significantly enhances modularity, reusability, and scalability within web applications, marking a new era in web development. This transformative technology has the potential to reshape both Component-Based Software Engineering (CBSE) and Design Systems, ensuring that software development becomes more efficient and design remains visually cohesive. By addressing the unique needs of web applications, customized built-in elements provide a path to a more streamlined and effective development process. Their adaptability allows for a high degree of customization and alignment with a brand's design language, fostering consistent user interfaces and experiences across various digital platforms. While the adoption of customized built-in elements brings immense promise, it also raises important considerations, such as browser compatibility, performance optimization, accessibility, security, and developer tooling. Addressing these issues is essential for the seamless and effective incorporation of these elements into web applications. Looking ahead, the future of customized built-in elements is bright, with potential applications in diverse fields and their integration into Progressive Web Apps (PWAs). However, ongoing efforts in standardization, accessibility, security, tooling, and community engagement will be crucial to unlock their full potential. In an ever-evolving digital landscape, HTML5 customized built-in elements serve as a beacon of innovation and a catalyst for enhanced web development practices. They empower developers to create efficient, adaptable, and user-centric web applications, ensuring that the web continues to evolve to meet the ever-growing expectations of digital consumers.

## AUTHORS

**Hardik Shah,** completed his MS in IT degree from Rochester Institute of Technology, New York, U.S.A. and has 12+ years of professional experience with UX Design and full-stack Web Development. Currently, he specializes in leading a Web UI development team building semantic and accessible Design System.

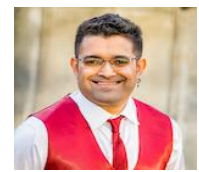